\documentclass[10pt,a4paper]{article}

\begin{document}

\title{OPERA and the GPS\\
{\normalsize\it A bon entendeur peu de mots suffisent}}
\author{Ll. Bel\thanks{e-mail:  wtpbedil@lg.ehu.es}}

\maketitle

\begin{abstract}

I comment about the adequacy of the GPS to model a particularly defined synchronization in a rotating frame of reference described in a general relativistic framework .

\end{abstract}

\vspace{1cm}

To decide wether an experiment contradicts a theory it is necessary to describe the experiment as a possible outcome in the framework of this theory. In this sense, in my opinion, the announcement of the OPERA team, \cite{OPERA} is not conclusive independently that it can be reproduced or not. And it will be until GPS specialists and the Relativity community may agree on the adequacy of the GPS to model a well defined synchronization described in a general relativistic framework.

I have written this short paper at the most simple level, hoping that it will be useful to those who are not familiar with the relativist aspect of the subject. (Second order corrections to the simple formulas here presented are included in \cite{Bel1}.)

\vspace{1cm}

Starting with Minkowski's line-element in a Galilean frame of reference using Cartesian coordinates:

\begin{equation}
\label{1}
ds^2=-c^2dt^2+d{\bar s}^2,
\end{equation}
with:

\begin{equation}
d{\bar s}^2=dx^2+dy^2+dz^2
\end{equation}
let us consider the time-like congruence:

\begin{equation}
t^\prime=t, \ x^\prime=x\cos\,\omega t-y\sin\,\omega t,
\ y^\prime=x\sin\,\omega t+y\cos\,\omega t, \ z^\prime=z.
\end{equation}
Inverting and substituting in (\ref{1}) we obtain, after dropping the primes, a new line-element that using a Weyl-like decomposition we write down as follows:

\begin{equation}
ds^2=-A^2(-cdt-f_1dx-f_2dy-f_3dz)^2+A^{-2}d{\bar s}^2,
\end{equation}
where now:
\begin{equation}
d{\bar s}^2=A^2d{\tilde s}^2
\end{equation}
with:

\begin{equation}
A=\sqrt{1-\frac{\omega^2\rho^2}{c^2}},\quad \rho=\sqrt{x^2+y^2}
\end{equation}
and:

\begin{equation}
f_1=A^{-2}\frac{\omega x}{c} ,  \ f_2=-A^{-2}\frac{\omega y}{c} ,  \ f_3=0
\end{equation}

Assuming that $\omega\rho/c$ is a small quantity and linearizing we get:

\begin{equation}
ds^2=-c^2dt^2+2\omega ydxdt-2\omega xdydt+\tilde {ds}^2
\end{equation}
where now:

\begin{equation}
d\tilde s^2=dx^2+dy^2+dz^2
\end{equation}
will be used as metric of reference\,\footnote{This concept is meant to give a meaning to the coordinates of space \cite{Bel2}} to calculate finite distances. That is all I need for the comments that I want to make.

A relativist has good arguments to claim that at this approximation he is using an appropriate setting: the line-element is stationary; the space coordinates $x,y,z$ are Cartesian harmonic coordinates and, and $t$ is also an harmonic coordinate since:

\begin{equation}
\frac{\partial f_1}{\partial x}+\frac{\partial f_2}{\partial y}+\frac{\partial f_3}{\partial z}=0
\end{equation}

The corresponding geodesic equations are:

\begin{eqnarray}
\frac{d^2x}{du^2}+2\omega\frac{dy}{du}\frac{dt}{du}=0 \\
\frac{d^2y}{du^2}-2\omega\frac{dx}{du}\frac{dt}{du}=0 \\
\frac{d^2z}{du^2}=0\\
\frac{d^2t}{du^2}=0
\end{eqnarray}
where $u$ is an affine parameter that runs, say, from $u=0$ to a sufficiently small value $u=u_f$. Let us consider an approximate solution for null geodesics:

\begin{eqnarray}
\label{16}
x=x_0+c_1u-c_2\omega u^2 \\
\label{17}
y=y_0+c_2u+c_1\omega u^2 \\
\label{18}
z=z_0+c_3u \\
\label{19}
t=u
\end{eqnarray}
with:

\begin{equation}
\label{20}
c_1^2+c_2^2+c_3^2 = c^2+2\omega(x_0c_2-2y_0c_1)
\end{equation}
that comes from requiring that the tangent vector at the origin to be a null vector.

Let us consider a light ray that starting from a location $P_0$ with coordinates $x_0,y_0,z_0$ at time $t=0$ reaches a location $P_1$ with coordinates $x_1,y_1,z_1$ at time $t=u_f$. This value of $u_f$ and the corresponding values of $c_1,c_2,c_3$  have to be calculated  using the four equations (\ref{16})-(\ref{18}) and (\ref{20}).

From (\ref{19}) it is obvious that the travel-time from $P_0$ to $P_1$ is $\Delta t=u_f$

On the other hand the distance from $P_0$ to $P_1$ is:

\begin{equation}
\Delta \tilde s=\sqrt{(x_1-x_0)^2+(y_1-y_0)^2+(z_1-z_0)^2}
\end{equation}
wherefrom it follows, at the assumed approximation, that:

\begin{equation}
\Delta \tilde s=\sqrt{c_1^2+c_2^2+c_3^2}u_f
\end{equation}
that taking into account (\ref{19}) leads to:

\begin{equation}
v_l\equiv \frac{\Delta \tilde s}{\Delta t} =c+\frac{\omega}{c}(c_2x_0-c_1y_0)
\end{equation}
a value that can be equal, greater or smaller than $c$. Notice however that the mean value of the speed of light in opposite directions is c. (This ceases to be true when terms of order $(\omega\rho/c)^2$ are taken in consideration).

Is this preceding result a violation of the postulate telling us that $c$, the coefficient of $dt^2$ in (\ref{1}), is a universal constant? Of course not. $c$ was measured  as the product of the frequency $\nu$ times the wave-length $\lambda$ of a particular cavity resonance \cite{Hall}, and the value that was obtained by no means depended on any relativistic global space-time model nor synchronization protocol. On the contrary, the formula above for $v_l$ depends crucially on both and on our choice to match them with a method of measuring distances and with our use of a clocks based distributed time. It turns out that now the GPS serves the two purposes and therefore it could be part of the problem if discrepancies between the theory and experiments are observed.

\vspace{1cm}

The conjecture that I want to make is that the GPS time may not be a good match to the formally defined synchronization described by the family of hyper-surfaces $t=const.$. This could be the case, for example, if it broke the spherical symmetry; a simple minded remark that follows from learning that the planes of the orbiting satellites are all inclined at $55^{\rm o}$ over the equator. In this case we could envision new relativistic models of time as for instance:

\begin{equation}
t^\prime=t+\frac{k}{2c^2}\omega(\rho^2-2z^2)
\end{equation}
where $k$ is a parameter to be determined by experiments of the OPERA type. This example is compatible with the axial symmetry and it is such that $t^\prime$ is also an harmonic coordinate. Checking the consistency of this choice, many experiments with a variety of initial and final conditions would be necessary to validate this choice, or any other of course.

\end{document}